# A Simultaneous-Movement Mobile Multiplayer Game Design Based on Adaptive Background Partitioning Technique

Samuel King Opoku, *Member, IEEE*

*Abstract*—Implementations of mobile games have become prevalent industrial technology due to the ubiquitous nature of mobile devices. However, simultaneous-movement multiplayer games – games that a player competes simultaneously with other players – are usually affected by such parameters as latency, type of game architecture and type of communication technology. This paper makes a review of the above parameters, considering the pros and cons of the various techniques used in addressing each parameter. It then goes ahead to propose an enhanced mechanism for dealing with packet delays based on partitioning the game background into grids. The proposed design is implemented and tested using Bluetooth and Wi-Fi communication technologies. The efficiency and effectiveness of the design are also analyzed.

*Index Terms*— Background partitioning, Communication technology, Game architecture, Latency, Mobile game, Multiplayer game, Online game, Packet delay, Simultaneous-movement game,

## I. INTRODUCTION

MOBILE games are video games played on mobile phones, smart phones, PDA or handheld devices and it is played using the technologies present on the device itself. A mobile game can be single or a multiplayer. In a multiplayer environment, the player collaborates or competes with other players who are playing the same game on their mobile device while connected via some network. Simultaneous-movement games are games that are based on turns. The participants play simultaneously and therefore waiting time is reduced [1]. These games are however affected by the game architecture and such network parameters as latency and the type communicate technology.

This paper reviews these important factors and the various mechanisms that have been proposed in solving these problematic parameters. It finally designs and implements an enhanced mechanism for addressing these parameters based on partitioning the game background into grids. The proposed design is implemented and tested using Bluetooth and Wi-Fi communication technologies such that mobile clients communicate with their servers through Bluetooth whereas inter-server communication is achieved through Wi-Fi. The efficiency and effectiveness of the design are also analyzed.

## II. LATENCY ISSUES

Latency has been the most crucial network parameter on line gaming. It refers to the time a terminal takes to send information to a server or to receive the reply. Interaction latency [2] is therefore defined as the time from the generation of interaction request till the appearance of updated image. Apparently, the interaction latency takes at least a roundtrip network transmission time between the mobile client and the rendering server. Latency varies greatly from network to network, wired networks to wireless network and as a function of congestion on a given network [3]. It can vary significantly from time to time and can kill the user experience in many latency-sensitive applications [2], [4]. Latency rate is usually affected by jitter and packet loss [3], [5]. Jitter refers to the delay variance which is usually caused by routers queuing packets because of congestion or prioritizing traffic. The margin of jitter is directly proportional to the margin of latency. A packet may not reach its destination due to network saturation, degradation of the signal through the network medium, faulty network or hardware and faulty packet [6]. A common way to reduce the effect of jitter is time stamping the packets. This allows the receiver to store the packets in a buffer until they are delivered in the right sequence with appropriate inter-arrival spacing based on the time stamps. [3], [7]

One technique for compensating latency is the use of image warping [8]-[11]. Image warping refers to the process of geometrically transforming two-dimensional picture or image. Although the word "Warp" may suggest radical distortion, the term "Image Warping" encompasses the whole range of transformation such as scaling or rotation to complex, irregular warp. The basic application of digital image warping is to change location, scale or orientation of the image. In warping technique, the rendering server generates a new image (from the previous image) and sends auxiliary information such as the coordinates of the new location of the generated image to mobile clients. The mobile client then generates images at the new viewpoint if any user interaction happens. A rendering server may produce two or more images as reference frames in order to reduce computational time. In such situations, Graphic Processing Unit (GPU) is usually used to accelerate search based algorithms [2], [11] in finding the required image to be displayed at the new viewpoint.

A proposed method for latency reduction is the Service Oriented Architecture (SOA) mostly in the form of web services with SOA backend [12], [13]. The mobile application usually calls many different SOA backend services to support the mobile process. However, performing a high number of network requests consumes a lot of energy on the mobile device. Another way to reduce perceived latency is by using rich client hybrid architecture [14], where the client can cache results locally. This architecture also allows some degree of off-line operations through the use of modern web technologies like Asynchronous JavaScript and XML (AJAX) [15]. Yet, using client-side cache to reduce latency perceived by the user is ideally suitable for non-mobile settings due to its poor usability issues and excessive power consumption [16]-[19]. An attempt of employing pre-fetching and caching that takes into account energy conservation has been described in [20]. These techniques assume a broadcast scenario, in which clients have to decide when to activate the network connection and actively pre-fetch data they observe from the broadcast channel. This is not applicable to mobile process participation, where a single client invokes several services in the backend SOA infrastructure. In [21] and [22], a general approach for caching web services for mobile devices that relies on rich JavaScript Client executed in the browser of the mobile device was proposed. The approach requires that the application developer specifies the pre-fetching and caching rules in great detail, instead of easing the burden of the developer.

An effective method [23] of handling latency in a highly interactive game is to adapt the server's update time T according to the heterogeneous latency. The protocol works as follow:

- The game status in the server gets updated every period, T.
- During that time, clients send their update information to the server
- Packets that arrive too late will however, be considered lost
- Clients are updated by the server

When latency occurs, the client suffers from missing game data. Dead reckoning is a way to estimate this missing data by taking into account recent positions, velocities and acceleration of game objects. Through the use of Dead Reckoning (DR) vector, a player sends his or her position in X, Y and Z coordinates, velocity, acceleration and other parameters such as pitch, roll and yaw to other players in the game sessions. When the vector is received by the other terminals or devices, they can predict the sender's future movement assuming that the velocity and the acceleration are unchanged. This prediction is done until the next vector is received which will give the receiver an updated state for the sender. When dead reckoning is only used to predict the client's position rather than frequently updating each peer with new information, bandwidth is conserved [24]. Dead reckoning, however, has some limitations. Its implementation means that all the clients have to run an algorithm to calculate the vectors while running the game [25] and this consumes processing power and battery's energy. They can slow down the client significantly. Dead reckoning is only useful when it is possible to predict a probable path for the game objects but if their predicted movements do not coincide with the client's actual movement, then the prediction wastes resources

## III. GAME ARCHITECTURE

There are three architectures for implementing mobile games [7]. These are peer-to-peer (P2P), client-server and network server. A P2P connection is only based on clients who are connected to each other. The game status is updated individually at every client [26]. This solution is perfect for the game provider [7], [26] because there is no need to invest in a server. Another advantage is that the network is very stable [7], [27] such that if one client goes down, the remaining clients still make up a network and continue to send information to each other. The drawbacks are related to security [26], [27]. All the clients have all the game information and it will be much easier for one gamer to cheat by hacking the game information in his terminal [3], [27]. To avoid divergence in game status due to delays, synchronization has to occur between clients [7], [26]. The network traffic generated increases exponentially as $n(n-1)$ where n is the number of clients and thus a player can easily run out of bandwidth [7], [27]. Fig. 1 illustrates a typical P2P architecture.

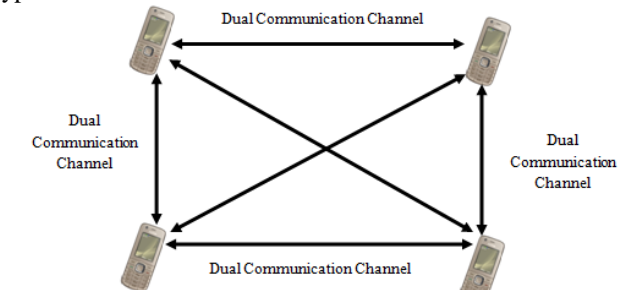

Fig. 1. A Typical P2P Architecture

A client-server connection, on the other hand, is based on a server. The server stores and processes all the game data it receives from all the connected clients. It only updates those clients with the data they need, thus every client receives a unique update [7]. The limited information that the server sends to its clients is good from traffic point of view and leads to lower latency [7], [28]. From a coding perspective, the model is preferable because little code needs to be added to support this sharing of state information and it can be easily be separated from the game code – a techniques suitable for multiplayer games and mobile games [3], [7]. A typical drawback is the high load at the server side. It is difficult in this architecture to have global knowledge of the game state [28] since the server updates each client individually within their scope and the data received by each client contains information about his or her scope only. Hence client-server model is better if the developer wishes to avoid players cheating in the game [7], [26]. The figure below illustrates the implementation of client-server architecture with four clients

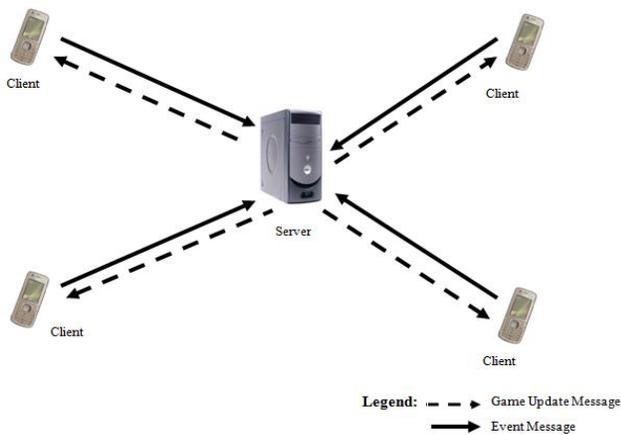

Fig. 2. Overview of Client-Server Architecture

The last architecture is the network server. This is a game genre with many participants connecting to the same server. The clients connect to one or more servers which are in turn interconnected with one another through a local network [7], [27]. The local network enables the servers to exchange a huge amount of data very quickly [7]. This model allows many clients to connect to a server without causing saturation of a single server [28], [29]. Fig. 3 is used to demonstrate the implementation of network-server architecture.

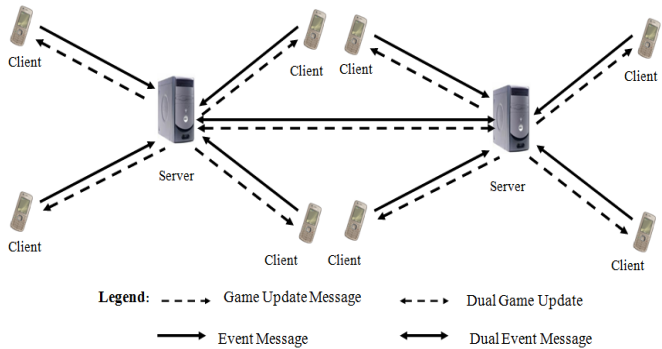

Fig. 3. Overview of Network-Server Architecture

## IV. MOBILE GAME COMMUNICATION TECHNOLOGIES

There must be a communication medium – be wired or wireless – for devices to communicate. Wireless technology is more convenient for mobile devices [30], [31] due to its flexibility in nature. For online mobile gaming, the various communication technologies employed are [3], [7], [28]: Bluetooth, General Packet Radio Service (GPRS), Universal Mobile Telecommunications Services (UMTS), Wide Code Division Multiple Access (WCDMA), and High-Speed Downlink Packet Access (HSDPA).

Bluetooth communication protocol has client-server architecture. The client initiates the connection and the server accepts or receives the connection [32]. Communication between devices depends on the type of data transferred. Object Exchange (OBEX) protocol supports exchange of such physical data as files and images. Logical Link Control and Adaptation Protocol (L2CAP) is used for sending packets between host and client whereas Radio Frequency COMMunication (RFCOMM) supports simple data transfer [32]-[34]. Although RFCOMM is easy to setup by providing Universally Unique Identifier (UUID) to indicate the service provided [35] yet a one-way communication link usually shuts down before data is transferred [32]. This problem is addressed by allowing bidirectional transmission before shutting down the link. This approach is not suitable from gaming perspective since it wastes bandwidth. Bluetooth is also limited by excessive power consumption [36], [37] and the power level of the energy source especially when the software development kit used for game development does not suite the handheld device [38], [39].

General Packet Radio Service (GPRS), adds packet data service to the Global System for Mobile (GSM) network. GPRS is the most widespread wide area wireless data service available [7], [40]. GPRS offers packet-based IP communication and builds upon the existing GSM technology and networks [41]. Enhanced Data GSM Environment (EDGE) is an enhancement of the GSM/GPRS network where the architecture is unchanged. By using different modulation scheme during the timeslot allocated for GPRS, the throughput can be increased [7], [40] and thus latency in EDGE is of a higher performance than that of GPRS. The advantage of EDGE is that if a packet fails to be transmitted, GPRS simply ignores it whereas EDGE will try to send it again using coding schemes with more error correction so that hopefully, the transmission will be successful [42]. Thus producing less packet loss makes it good communication technology from gaming perspective.

UMTS is a packet switching 3G technology which in comparison to EDGE, offers significantly greater data transfer and allows simultaneous voice/data communication [42], [43] which is suitable from a cellular online gaming point of view as the gaming session will not be interrupted by an incoming call. These terminals are however, big, heavy and battery consuming [7], [43] although the bigger screen gives an advantage towards gaming.

WCDMA is built on Code Division Multiple Access signaling method [42]. It has the advantage of offering soft handover between cells [43]. Thus less burst traffic can be expected which is better from gaming point of view. However, during hard handover when the terminal does not have any contact with the base station, burst traffic can occur since the terminal cannot transfer any data.

HSDPA offers higher bit rates and lower round trip delay [43] enabling applications such as multi-user gaming. HSDPA is an improvement of UMTS, built on the implementation of a new WCDMA channel [43], [44]. A drawback is that HSDPA geographical coverage is limited to big cities [7], [44].

## V. BACKGROUND PARTITIONING DESIGN

The background partitioning design is basically used to
- assign coordinate system to game ground so that the game parameters can be accordingly controlled
- control memory and processing power
- predict the subsequent game action when game packets are lost or when the latency rate is very high

## A. Assigning Coordinate System

Consider a 3-Dimensional background (x, y, z) which is shown in the figure below:

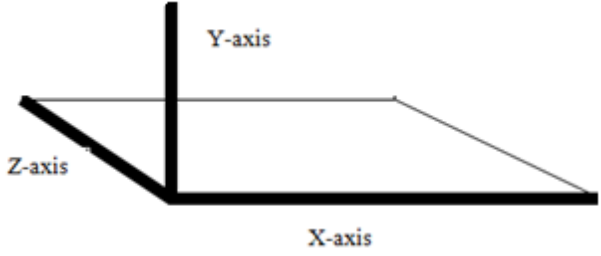

Fig. 4. 3-D Architecture of a Game Ground

The background partitioning algorithm ignores the y-axis values and then considers (x, z) as a 2-Dimensional background. The (x, z) axes are partitioned as shown below:

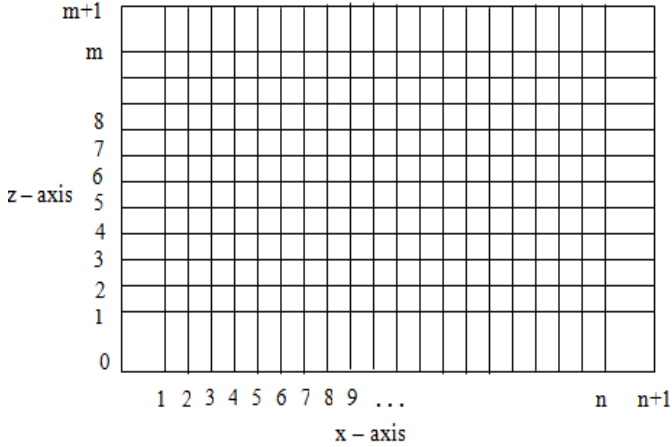

Fig. 5. Background Partitioning Design

The points with the following coordinates: (0, 0), (n+1, 0), (n+1, m+1) and (0, m+1) serve as the boundary of the entire game system whereas the playing ground boundary is characterized by the points with the following coordinates (1, 1), (n, 1), (n, m) and (1, m). The difference between two adjacent points such as (n+1) – n form the background partitioning point interval denoted by $I$. The partitioning system is used to determine the x and z coordinates of every player of the game. The partitioning is based on
- Game level which determines the acceleration of the game and the players
- Interaction latency rate which corresponds to the delay variance between sending and receiving game packets

Thus the game partitioning parameter, ρ, is a function of latency rate, L and the game level, G. The higher the latency rate, the greater the delay and hence the smaller the intervals between game ground points. However, the higher the game level, the higher the intervals between ground points. It, therefore, implies that $\rho \propto \frac{G}{L}$ and hence $\rho = \theta \frac{G}{L}$. θ is the partitioning parameter constant which depends on the type of communication technology and the type of game architecture.

To minimize computational overload, partitioning parameter, ρ, for both x and z axes are equal. Thus both axes have the same scale for the coordinate system. The background partitioning point interval, $I$, is computed using the partitioning parameter, ρ, and the width of the screen. The basic condition is that there should be at least $R$ points on both x and z axes depending on the type of game implemented. Let $w_s$, be the width of the screen. Thus if $\rho \leq w_s/R$ then $I = \rho$. However, if $\rho > w_s/R$ but $\rho \leq (2*w_s)/R$ then $I = \rho/2$. On the other hand, if $\rho > (2*w_s)/R$, $I = w_s/R$. Using the background partitioning interval, $I$, the coordinates of x and z are computed as $x_{n+1} = x_n + I \leq w_s$ and $z_{n+1} = z_n + I \leq w_s$ given that $n \in \mathbb{Z}^+$.

Let (x, z) be the current position of a game player. The next position is determined as $(x \pm I, z \pm I)$ for ordinary players and $(x \pm \varphi I, z \pm \varphi I)$ for "*specially-talented*" players with the ball. The speed of the ball is determined by the player handling it. φ represents the number of intervals that needs to be combined so that the speed of the player is doubled or tripled. Thus $\varphi \in \{2,3\}$ with $\varphi = 3$ used for "*world class*" players. A "*specially-talented*" player without the ball has its next position determined as $(x \pm I, z \pm I)$.

The background partitioning parameter is adaptive in that the parameter is computed in regular intervals and when there is a change in the value of ρ, the background is re-partitioned using the new value of ρ. From $\rho = \theta \frac{G}{L}$, it implies that when the latency rate is very high, the background partitioning point interval diminishes allowing the game players and the object of concern to move slowly. Thus, $\lim_{L \to \infty} \rho \cong 0$ indicating that the game players and the object of concern turn to be motionless when the network connection increasingly becomes slower.

## B. Managing Memory and Processing Power

The partitioning mechanism divides the game ground into regions. Each four intercepting lines create a region. In order to manage resource such as memory and processing power efficiently, the regions are categorized into More Detailed Region (MDR) and Less Detailed Region (LDR).

In such simultaneous-movement multiplayer games as football, basketball and hockey, the region surrounding the object of concern – the ball – is the MDR whereas the other regions constitute the LDR.

To determine the boundary of the MDR, let the current position of the ball be $(\propto, \beta)$ and $I$ be the interval between points on the game ground. The coordinates of the MDR is $(\propto \pm \mu I, \beta \pm \mu I)$ where $\mu \geq 1$ is a constant – called game region constant – that depends on the size of the memory and the processing power. Also $\forall \propto \pm \mu I$ and $\beta \pm \mu I$ the following conditions should be satisfied: $0 \leq \propto \pm \mu I \leq n + 1$ and $0 \leq \beta \pm \mu I \leq m + 1$. Thus $\mu I$ gives the number of the intervals required to add to or deduct from the coordinate of the ball to obtain the MDR. It therefore, follows that the coordinate of the boundary of MDR are: $(\propto - \mu I, \beta - \mu I)$, $(\propto - \mu I, \beta + \mu I)$, $(\propto + \mu I, \beta - \mu I)$ and $(\propto + \mu I, \beta + \mu I)$.

If there exists a coordinate such that $\propto + \mu I > n + 1$ or $\propto - \mu I < 0$ then $\propto + \mu I = n + 1$ and $\propto - \mu I = 0$. Similarly, If there exists a coordinate such that $\beta + \mu I > m + 1$ or $\beta - \mu I < 0$ then $\beta + \mu I = m + 1$ and $\beta - \mu I = 0$.

To illustrate, consider a game ground with the background partitioning point interval, $I = 1$ and the game region constant, $\mu = 1$. The MDR is shown in the figure below:

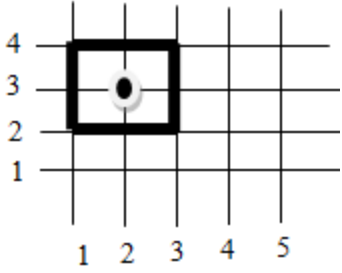

Fig. 6. Illustration of More Detailed Region

The object under consideration, the ball, is at point (2, 3). With I = 1 and μ = 1 the MDR is described by the following points using ($\propto \pm \mu I, \beta \pm \mu I$): (1, 2), (1, 4), (3, 2) and (3, 4).

The Less Detailed Region (LDR) does not appear on the screen and thus the memory that can be used to store detailed information about the LDR is freed. All the players and activities in the LDR are freeze. Once a ball moves to the LDR, a new MDR is assigned and the previous MDR is frozen and treated as LDR.

The y-axis values are designed so that memory and processing power are effectively controlled. An interview with eighty computer game players demonstrated that they do not take notice of the height of game players into consideration when playing games like football and hockey. In view of that all the game players are designed with the same height and hence they all occupy the same y-axis value of the game background. The point interval for y-axis denoted by $I_y$ is given by $I_y = (Y_{max} - Height_{player})/3$. A player can jump from $Height_{player}$ to $Height_{player} + I_y$ whereas the object of concern (usually the ball) can move to a maximum height of $Height_{player} + 2 * I_y$. However, in a basketball game, a player can jump from $Height_{player}$ to $Height_{player} + 2 * I_y$

### C. Managing Packet Loss and Excessive Packet Delay

Game data is lost when packets are lost. In excessive packet delay or occurrence of very high latency rate, game packets are considered to be lost. The mechanism employed in managing packet loss divides the game progress into game states over time spent. A round trip communication between the online players indicates a complete state of the game.

Let $S_n$, $n \in \mathbb{Z}^+$ denotes the game states such that if $S_k$ is the current state, $\forall k \in n$ and $k \geq 1$, the immediate previous state is $S_{k-1}$. Also let T be the time spent, in seconds, in the game from $S_0$ to $S_k$, it follows that $T(S_k) > T(S_{k-1}) > 0$. When $T = 0$, that is at the beginning of the game, $S_0$ is set to the default setting. The default setting for a typical such multiplayer game as football is the default positions of the players and the ball when the game is about to start. The MDR is in the center of the playing ground.

The mechanism employed is applied from the second round trip communication onwards. This mechanism also sets the background partitioning parameter, game acceleration and the players' velocities of $S_n$ to be equal to that of $S_{n-1}$ $\forall n > 1$ where $S_n$ is the state whose game data is missing and the game parameters need to be computed. The background partitioning point interval, $I$, for $S_{n-1}$ is therefore used to determine the positions of players and the ball in the current state, $S_n$.

The set of parameters needed for the current state, $S_n$, is given by
$$S_n = \{I_{n-1}, P^k{}_n, B_n\}$$
where
$I_{n-1}$ is the background partitioning point interval
$P^k{}_n = (X^k{}_n, Y^k{}_n, Z^k{}_n)$ is the coordinate of the players such that $\forall k$, $k \in \mathbb{Z}^+ \leq q$ represents the $k^{th}$ player out of the total $q$ players in the MDR. When $k = 0$, there is no player in the MDR.
$B_n = (X_n, Y_n, Z_n)$ is the coordinate of the ball

The following are general parameters used to determine the coordinates of the ball and the players.
Let
$I$ be the background partitioning point interval
$X_{max}$ be the x-axis point limit of the entire game system
$Y_{max}$ be the y-axis point limit of the entire game system
$Z_{max}$ be the z-axis point limit of the entire game system
$I_y$ be the interval between the y-axis values

The design for obtaining player's coordinate assumes that a player can only move in one direction at any given time and thus for any particular time, only one of the three axes is affected leaving the other two axes unchanged.

Let
$X^k{}_p$ be the x-axis value of the $k^{th}$ player in the current state
$Y^k{}_p$ be the x-axis value of the $k^{th}$ player in the current state
$Z^k{}_p$ be the x-axis value of the $k^{th}$ player in the current state

The algorithm below is used to determine whether the player is stationary or not:
If ($|X^k{}_{p-1} - X^k{}_{p-2}| < I$ and $|Y^k{}_{p-1} - Y^k{}_{p-2}| < I_y$ and $|Z^k{}_{p-1} - Z^k{}_{p-2}| < I$)
Then the player is assumed to be motionless and it continues to be at its position. Thus
$$X^k{}_p = X^k{}_{p-1}, Y^k{}_p = Y^k{}_{p-1} \text{ and } Z^k{}_p = Z^k{}_{p-1}$$

To determine the x-coordinate, the following algorithm is used:
If ($|X^k{}_{p-1} - X^k{}_{p-2}| \geq I$ and $|Y^k{}_{p-1} - Y^k{}_{p-2}| < I_y$ and $|Z^k{}_{p-1} - Z^k{}_{p-2}| < I$)
Then
$$0 \leq [X^k{}_p = X^k{}_{p-1} + (sign \; of \; X^k{}_{p-1}) * I] \leq (X_{max} - I)$$

Similarly, the z-axis coordinate is determined by the algorithm:
If ($|X^k{}_{p-1} - X^k{}_{p-2}| < I$ and $|Y^k{}_{p-1} - Y^k{}_{p-2}| < I_y$ and $|Z^k{}_{p-1} - Z^k{}_{p-2}| \geq I$)
Then
$$0 \leq [Z^k{}_p = Z^k{}_{p-1} + (sign \; of \; Z^k{}_{p-1}) * I] \leq (Z_{max} - I)$$

The y-axis is designed to have two states in managing packet loss mechanism. A player may jump to increase y-axis value or descend from a height after jumping to decrease the y-axis value. Thus

If $(|X^k_{p-1} - X^k_{p-2}| < I$ and $Y^k_{p-1} - Y^k_{p-2} \geq I_y$ and $|Z^k_{p-1} - Z^k_{p-2}| < I)$

Then the player jumped and has to come down. Thus
$X^k_p = X^k_{p-1}, Y^k_p = Y^k_{p-2}$ and $Z^k_p = Z^k_{p-1}$

Else If $(|X^k_{p-1} - X^k_{p-2}| < I$ and $Y^k_{p-1} - Y^k_{p-2} < I_y$ and $|Z^k_{p-1} - Z^k_{p-2}| < I)$

Then the player is down or came down from jumping and has to retain its previous value. Thus
$X^k_p = X^k_{p-1}, Y^k_p = Y^k_{p-1}$ and $Z^k_p = Z^k_{p-1}$

The ball alone can move in one direction along x-axis or z-axis only. Thus for any given game state, only one of the two axes (x and z) is affected leaving the other unchanged. However, a ball can move along x-axis direction altering the y-axis accordingly at the same time or it can move along z-axis direction whereas the y-axis is changed accordingly at the same time. This depends on the game control key(s) used in $S_{n-1}$ state.

Let
$X_b$ be the x-axis coordinate of the ball in the current state
$Y_b$ be the x-axis coordinate of the ball in the current state
$Z_b$ be the x-axis coordinate of the ball in the current state
$\tau$ be the strength of the game control key pressed by the online player such that $\forall \tau, \tau = 0, 1, 2$

$\tau = 0$ when the control key suggests that the value of the y-axis coordinate has to be retained and thus $Y_b = Y_{b-1}$. The design assumes that from any state $S_{n-1}$ to $S_n$, the ball cannot move beyond two consecutive interval points of the y-axis and hence $0 \leq \tau \leq 2$. Increasing $\tau$ indicates that the ball ascends along the y-axis such that $Y_{max} \geq Y_b \geq Y_{b-1} + \tau * I_y$ and $\tau = \tau + 1 < 2$ whereas decreasing $\tau$ indicates that the ball descends so that $0 \leq Y_b \leq Y_{b-1} + \tau * I_y$ and $\tau = \tau - 1 > 0$.

The ball is assumed to be motionless when the condition shown below is satisfied:
If $(X_{b-1} == X_{b-2}$ and $Y_{b-1} == Y_{b-2}$ and $Z_{b-1} == Z_{b-2})$
Then
$X_b = X_{b-1}, Y_b = Y_{b-1}$ and $Z_b = Z_{b-1}$

The algorithm below computes the x-coordinate of the ball in the current state given that y-coordinate can change:
If $(|X_{b-1} - X_{b-2}| < I$ and $Z_{b-1} == Z_{b-2})$ Then
$0 \leq [X_b = X_{b-1} + (sign\ of\ X_{b-1}) * I] \leq X_{max}$
and
$0 \leq [Y_b = Y_{b-1} + \tau * I_y] \leq Y_{max}$

Similarly, the z-coordinate of the ball is computed given that the y-coordinate can change:
If $(X_{b-1} == X_{b-2}$ and $|Z_{b-1} - Z_{b-2}| < I)$ Then
$0 \leq [Z_b = Z_{b-1} + (sign\ of\ Z_{b-1}) * I] \leq Z_{max}$
and
$0 \leq [Y_b = Y_{b-1} + \tau * I_y] \leq Y_{max}$

## VI. TESTING AND ANALYSIS OF THE PROPOSED DESIGN

### A. Architecture of the Testing System

The testing system is made up of two mobile phones and two computer systems. Each phone is connected to a computer system through Bluetooth. The computer systems are connected via Wi-Fi. The computer systems act as servers to the mobile phone and thus the system implements network server game architecture. This architecture allows manual manipulation at the server side regarding packet delays. The general overview of the system is illustrated in the figure below:

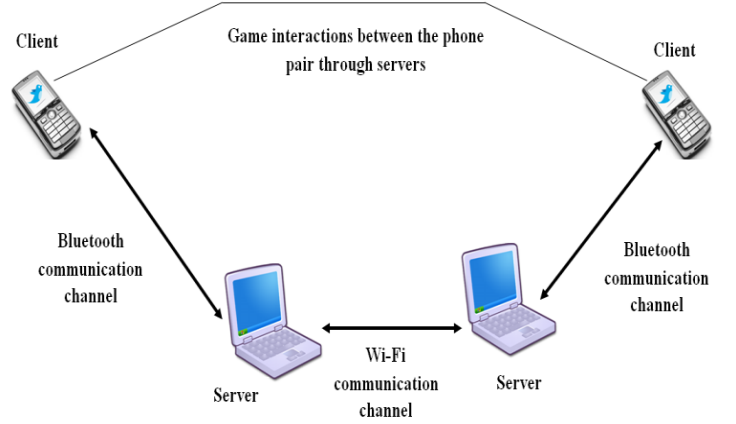

Fig.7. Overview of the Implementation Architecture

For the servers to be connected, one of the servers initiates a connection and the other server accepts. Thus, there is a client-server relationship between the servers. The flowchart below illustrates the connection process of the testing system.

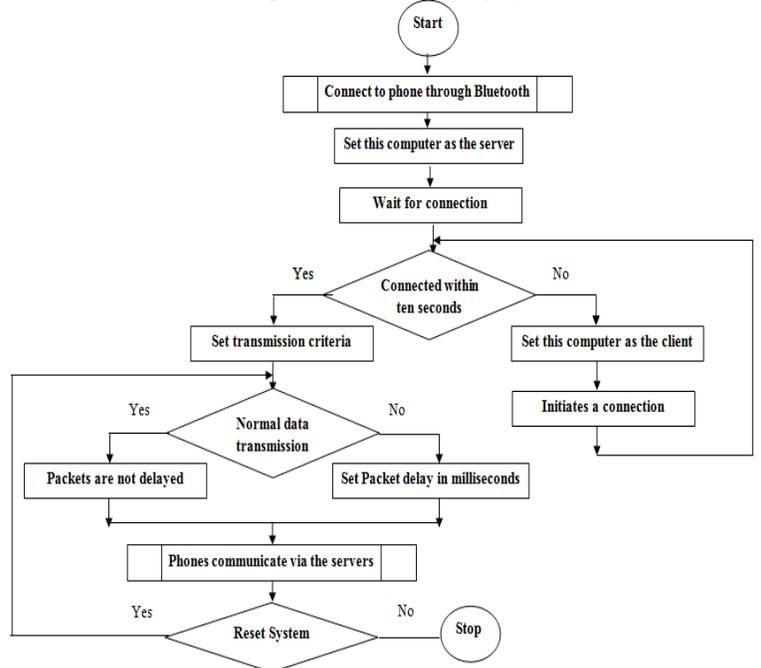

Fig.8. Connection Process of the Testing System

## B. Testing of the Implemented System

Soccer was the game simulated using the above architecture. In a displayed view, a user will only see what is demonstrated in Fig. 9. The figure below represents MDR with only one player near the throw line.

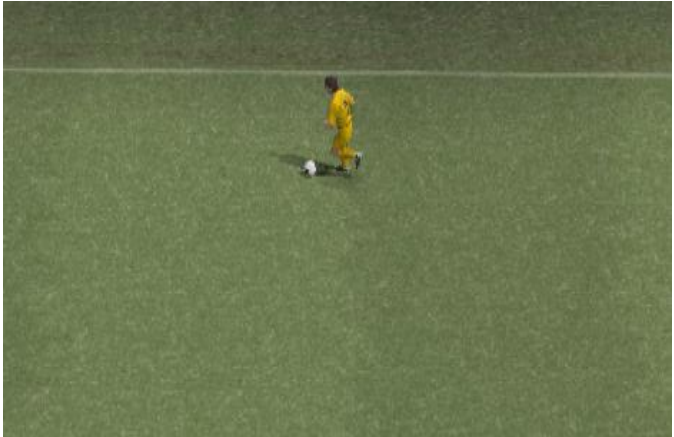
Fig.9. Display View of a Game Ground

However, beneath the display view is the implementation of the partitioning system. The figure below demonstrates the implementation of the background partitioning system. The ball occupies the point (0.64, 0.92) whereas the player is at (0.66, 0.92).

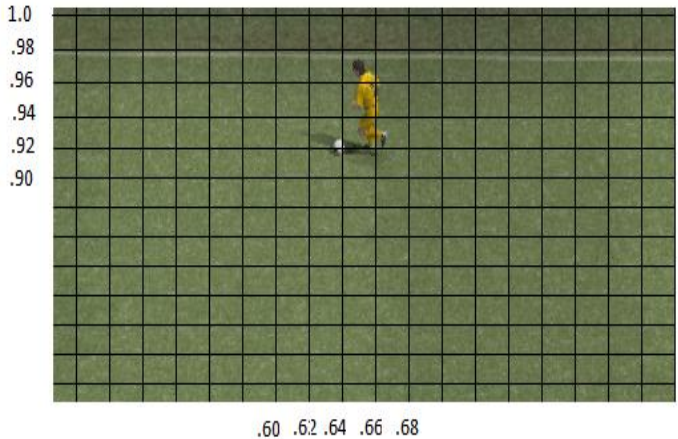
Fig.10. Background Partitioning System of the Game Ground

The above partitioning system is implemented in a worse situation with higher game level and very high latency rate. The point interval is very small ($I = 0.02$) and thus the player with the ball has different coordinate values from that of the ball. However, the above situation can also arise when the game is played with the least game level and a tolerable latency rate. The next position of the player from Fig. 10 can be one of the following points (0.66, 0.92), (0.66, 0.90), (0.66, 0.94), (0.64, 0.92) and (0.68, 0.92)

The figure below demonstrates a game played with higher game level with a normal latency rate. In such case, the player with the ball has the same coordinate as the ball. From the figure below, the player and the ball occupy the point (6.0, 5.5). This situation reduces memory consumption and also controls processing power. The other player approximately occupies the point (5.0, 4.5). The approximate points are used in predicting the subsequent coordinates when game data is lost.

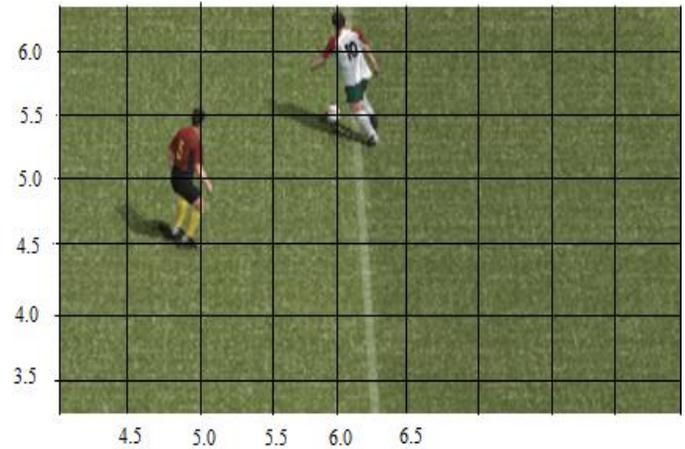
Fig.11. Normal Background Partitioning System of the Game Ground

## C. Analysis of the Implemented System

Unlike various game designs that rely on the velocity of the game players and the acceleration of the game to determine the subsequent actions of game objects, this design ignores players' velocity. The coordinate of the game objects can easily be determined using the background partitioning system. Using two previous states of the game, the coordinates of the game objects of the next state are effectively determined.

The major drawback of the design is its inability to resist frequent latency fluctuations. This inevitably kills user's interest. This happens as the background partitioning also adjusts frequently to reflect the latency rates. Users therefore will find it difficult to predict the outcome of their actions.

## VII. CONCLUSION

In this paper, the researcher reviewed the various factors that affect online multiplayer game which is played simultaneously. The researcher went ahead to propose a robust design based on background partitioning technique that has the ability to adapt to different latency. The work serves as a basis for developing online game systems that eliminate excessive coding which usually takes into consideration parameters such as velocity, acceleration, pitch, roll and yaw of game players.

The efficiency and the effectiveness of the design were tested and analyzed. The implemented system has the following characteristics:
- It does not require many parameters in its computations
- It conserves memory and processing power by minimizing computation

REFERENCES

[1] Nokia White Paper, *"Overview of Multiplayer Mobile Game Design"*, Available: http://www.forum.nokia.com
[2] S. Shi, "*Reduce latency: The Key to Successful Interactive Remote Rendering Systems*", IEEE International Conference on Pervasive Computing and Communications Workshops, pp. 391-392, 2011
[3] A. Spurling, "*QoS Issues for Multiplayer Gaming*", Case study at Cardiff university 2004, http://users.cs.cf.ac.uk/O.F.Rana/data-comms/gaming.pdf
[4] M. Claypool, K. Claypool, *"Latency Can Kill: Precision and Deadline in Online Games"*, Proceedings of the first annual ACM SIGMM conference on Multimedia systems, New York, NY, USA, pp. 215–222, 2010
[5] R. Rakesh, N. Amiya, "*An Efficient Video Adaptation Scheme for SVC Transport Over LTE Networks*", IEEE 17th International Conference on Parallel and Distributed Systems (ICPADS), pp. 127-133, 2011
[6] O.J.S. Parra, A. P. Rios, G. Lopez Rubio, "*Quality of Service Over IPv6 and IPv4*", 7th International Conference on Wireless Communications, Networking and Mobile Computing (WiCOM), pp. 1-4, 2011
[7] C. Westermark, *"Mobile Multiplayer Gaming"*, Master Thesis, School of Information and Communication Technology, Royal Institute of Technology, web.it.kth.se, pp. 20-45, 2007
[8] W. R. Mark, G. Bishop, L. McMillan, *"Post-Rendering Image Warping for Latency Compensation,"* Chapel Hill, NC, USA, Tech. Rep., 1996
[9] S. Shi, M. Kamali, J. C. Hart, K. Nahrstedt, R. H. Campbell, *"A High-Quality Low-Delay Remote Rendering System for 3D Video"*, Proceedings of the eighteen ACM international conference on Multimedia, New York, NY, USA, 2010
[10] S. Shi, W. J. Jeon, K. Nahrstedt, R. H. Campbell, *"Real-time Remote Rendering of 3D Video for Mobile Devices"*, Proceedings of the seventeen ACM international conference on Multimedia. New York, NY, USA, pp. 391-400, 2009
[11] W. Yoo, S. Shi, W. J. Jeon, K. Nahrstedt, R. H. Campbell, *"Real-time Parallel Remote Rendering for Mobile Devices using Graphics Processing Units"*, in ICME, IEEE, pp. 902–907, 2010
[12] M. Muhlhauser, I. Gurevych, *"Ubiquitous Computing Technology for Real Time Enterprises"*, Information Science Reference, 2008.
[13] R. Tergujeff, J. Haajanen, J. Leppanen, S. Toivonen, *"Mobile SOA: Service Orientation on Lightweight Mobile Devices"*, in ICWS, IEEE, pp. 1224–1225, 2007
[14] V. Gruhn, A. Kohler, *"Aligning Software Architectures of Mobile Applications on Business Requirements,"* in WISM, 2006.
[15] L. Hamdi, H. Wu, S. Dagtas, and A. Benharref, *"Ajax for Mobility: MobileWeaver AJAX Framework"*, Proceedings of the WWW. ACM, pp. 1077–1078, 2008
[16] M. Pervila, J. Kangasharju, *"Performance of Ajax on Mobile Devices: A Snapshot of Current Progress"*, 2nd International Workshop on Improved Mobile User Experience, 2008.
[17] V. N. Padmanabhan, J. C. Mogul, *"Using Predictive Pre-fetching to Improve World Wide Web Latency"*, ACM SIGCOMM Review, Vol. 26, No. 3, pp. 22–36, 1996.
[18] L. Fan, P. Cao, W. Lin, Q. Jacobson, *"Web Pre-fetching Between Low-Bandwidth Clients and Proxies: Potential and Performance"*, in ACM SIGMETRICS, pp. 178–187, 1999
[19] A. N. Eden, B. W. Joh, and T. Mudge, *"Web Latency Reduction via Client-Side Pre-fetching"*, in IEEE ISPASS, pp. 193–200, 2000
[20] G. Cao, *"Proactive Power-Aware Cache Management for Mobile Computing Systems"*, IEEE Transactions on Computers, Vol. 51, No. 6, pp. 608–621, 2002.
[21] K. Elbashir, R. Deters, *"Transparent Caching for Nomadic WS Clients"*, in IEEE ICWS, pp. 177–184, 2005
[22] D. Schreiber, E. Aitenbichler, A. Goeb, M. Muhlhauser, "*Reducing User Perceived Latency in Mobile Processes*", IEEE International Conference on Web Services (ICWS), pp. 235-242, 2010
[23] Wu-chang, F Feng, *"Provisioning On-line Games: A Traffic Analysis of a Busy Counter-Strike Server*", 2002, http://www.imconf.net/imw-2002/imw2002-papers/168.pdf
[24] S. Aggarwal, *"Accuracy in Dead-Reckoning Based Distributed Multi-Player Games"*, 2004, Available: http://www.sigcomm.org/sigcomm2004/workshop_papers/net610-aggarwal.pdf
[25] W. Cai, *"An Auto-adaptive Dead Reckoning Algorithm for Distrubuted Interactive Simulation"*, Proceedings of the thirteenth workshop on Parallel and distributed simulation, Atlanta, Georgia, United States, pp. 82-89, 1999
[26] T. Moser, "*Design and Implementation of a Multiplayer Peer-to-Peer Game for Android Mobile Devices*", Master Thesis, Department of Informatics, University of Zurich, pp. 25-40, 2010
[27] R. Schollmeier, "*A Definition of Peer-to-Peer Networking for the Classification of Peer-to-Peer Architectures and Applications*", Peer-to-Peer Computing, pp. 101–102, 2001.
[28] C. Xin, "*Multiplayer Game in Mobile Phone Serious Game*", International Joint Conference on Artificial Intelligence, JCAI '09, pp. 56-58, 2009. .
[29] S. M. Malfatti, F. Ferreira dos Santos, S. Rodrigues dos Santos, "*Using Mobile Phones to Control Desktop Multiplayer Games*", Brazilian Symposium on Games and Digital Entertainment (SBGAMES), pp. 230-238, 2010
[30] Z. Li-zhong, Y. Zheng-lin, G. Hai-feng, "*Study on Simulation Technology of 3G Mobile Communication Network*", WRI Global Congress on Intelligent Systems, GCIS '09.Volume: 3, pp. 236-241, 2009
[31] H. Wu, "*Some Thoughts on the Transformation of Information and Communication Technologies*", IEEE Technology Time Machine Symposium on Technologies Beyond 2020 (TTM), pp. 1, 2011
[32] S. Rathi, "*Infrastructure: Bluetooth Protocol Architecture*", Microware Architect, Microware System Corporation, p. 1 – 6, 2000
[33] B. Hopkins and R. Antony, *"Bluetooth for Java"*, ISBN 1-59059-78-3, pp 33-35, 2003.
[34] A. Zapater, K. Kyamakya, S. Feldmann, M. Kruger, I. Adusei, *"Development and Implementation of a Bluetooth Networking Infrastructure for a Notebook"*, University Scenario, International Conference on Wireless Networks, 2003
[35] S. K. Opoku, "*An Indoor Tracking System Based on Bluetooth Technology*", Cyber Journals: Multidisciplinary Journals in Science and Technology, Journal of Selected Areas in Telecommunications (JSAT), Vol. 2, No. 12, pp. 1-8, December Edition, 2011
[36] Bluetooth Special Interest Group Specification, "*Specification of the Bluetooth System Core*" Vol 1 and Vol 2, Versions 1.1, February 22, 2001
[37] M. M. Organero, S. K. Opoku, "*Using NFC Technology for Fast-Tracking Large-Size Multi-Touch Screens*", Cyber Journals: Multidisciplinary Journals in Science and Technology, Journal of Selected Areas in Telecommunications (JSAT), Vol. 2, No. 4, pp. 65-70 April Edition, 2011
[38] S. K. Opoku, "*Performance Enhancement of Large-Size NFC Multi-Touch System*", Cyber Journals: Multidisciplinary Journals in Science and Technology, Journal of Selected Areas in Telecommunications (JSAT), Vol. 2, No. 10, pp. 52-57, October Edition, 2011
[39] S. K. Opoku, *"Exploring Near Field Communication Multi-Touch"*, Master Thesis, Telematics Engineering Department, Universidad Carlos III de Madrid, pp. 1-52, 2011.
[40] Usha Communications Technology, "*White Paper of GPRS*", 26 June 2000
[41] B. Ghribi, L. Logrippo, *"Understanding GPRS: The GSM Packet Radio Service"*, School of Information Technology and Engineering, University of Ottawa, 2009.
[42] V. Sami, K. Katja, *"Positioning Edge in the Mobile Network Evolution"*, Master Thesis, Helsinki University of Technology, pp. 25-40, 2007.
[43] P. Rysavy, "*EDGE, HSDPA and LTE*", Rysavy Research Developed for 3G America, September 2006
[44] M. Akervik, "*Network Gaming: Performance and Traffic Modeling*", Master Thesis, Royal Institute of Technology (KTH), Department of Communication systems, November 2006.
[45] Amplified Engineering, *"Delivering Optimized Solution"*, 2004, http://www.amplified.com.au/Application_Ethernet_Device_to_Server_Internet_SMS.aspx